\documentclass[12pt,preprint]{aastex}

\usepackage{graphicx}        
\usepackage{natbib}         
\usepackage{amssymb}        
\usepackage{amsmath}
\usepackage{color}           
\usepackage{url}             
\usepackage{epstopdf}

\shorttitle{Double-diffusive mixing in stellar interiors in the presence of horizontal gradients}
\shortauthors{Medrano, Garaud, \& Stellmach}

\newcommand{\be}{{\bf e}}
\newcommand{\bu}{{\bf u}}

\begin{document}


\title{Double-diffusive mixing in stellar interiors in the presence of horizontal gradients}


\author{M. Medrano$^1$, P. Garaud$^1$ \& S. Stellmach$^2$}
\affil{$^1$ Department of Applied Mathematics and Statistics, Baskin School of Engineering, University of California Santa Cruz, 1156 High Street, Santa Cruz CA 95060 \\ 
$^2$ Institut f\"ur Geophysik, Westf\"alische Wilhelms-Universit\"at
M\"unster, M\"unster D-48149, Germany 
 }


\begin{abstract}
We have identified an important source of mixing in stellar radiation zones, that would arise whenever two conditions are satisfied: (1) the presence of an inverse vertical compositional gradient, and (2) the presence of density-compensating horizontal gradients of temperature (alternatively, entropy) and composition. The former can be caused naturally by any off-center burning process, by atomic diffusion, or by surface accretion. The latter could be caused by rotation, tides, meridional flows, etc. The linear instability and its nonlinear development have been well-studied in the oceanographic context. It is known to drive the formation of stacks of fingering layers separated by diffusive interfaces, called intrusions. Using 3D numerical simulations of the process in the astrophysically-relevant region of parameter space, we find similar results, and demonstrate that the material transport in the intrusive regime can be highly enhanced compared with pure diffusion, even in systems which would otherwise be {\it stable} to fingering (thermohaline) convection. 
\end{abstract}





\section{Introduction}

Double-diffusive instabilities have long been known to drive mixing in stars.
\citet{schwartzchildharm1958} and \citet{kato66} first considered the combined effects of an unstable entropy gradient and a stable compositional gradient ($\mu-$gradient, hereafter) to 
study what is commonly known as ``semi-convection". Meanwhile, \citet{ulrich1972} and \citet{kippenhahn80} investigated the opposite scenario of a stable entropy gradient with an unstable $\mu-$gradient, known as ``thermohaline" or ``fingering" convection. In both cases, the overall density gradient is stable to dynamical convection (so the system is Ledoux-stable) but the rapid diffusion of heat compared with the slower molecular diffusion of composition can destabilize the fluid, and trigger some level of turbulent mixing \citep[see the reviews by][for instance]{Garaud13,Radko13}. 

For the next few decades, however, the lack of experimental evidence prevented astrophysicists from estimating how much mixing these instabilities could generate. Only recently has it become possible to follow their nonlinear development numerically, and quantify the induced turbulent fluxes. Much progress has been made in this respect thanks to three-dimensional (3D) direct numerical simulations (DNS), both in the case of semi-convection \citep{Mirouhal12,Woodal13} and of fingering convection \citep{Traxleral2011,Brownal2013}. In this work, we focus on the latter. 

\citet{Brownal2013} proposed a parameter-free, experimentally calibrated model for mixing by fingering convection in non-rotating stellar interiors, that has been successfully tested against 3D DNS. The only limitation of this model is that it assumes a purely one-dimensional background, with both entropy- and $\mu$-gradients aligned with gravity. This assumption is consistent with the assumptions of most 1D stellar models, but does not necessarily represent actual conditions in stellar interiors. Indeed, a number of situations may arise in which weak horizontal gradients are present. As we demonstrate here, these can have significant impacts on mixing by double-diffusive instabilities. 

Horizontal gradients are expected whenever a star is rotating. For instance, rotation affects the convective efficiency differently at different latitudes, presumably causing pole-to-equator entropy gradients in adjacent radiative regions. It also generates large-scale meridional flows, which burrow into the nearby radiative region and drive further inhomogeneities \citep[e.g.][]{SpiegelZahn1992,WoodBrummell2012}. In radiation zones, the centrifugal force causes a misalignment between isotherms and isobars, which also break the spherical symmetry -- an effect at the source of Eddington-Sweet flows. Even if the star is rotating too slowly for this misalignment to be significant today, its spin-down history will have generated radial differential rotation, which is always accompanied by large-scale meridional flows, and therefore by lateral entropy gradients  \citep{Zahn1992,OglethorpeGaraud2013}. Finally, interactions with binary companions cause tides, which are also sources of asymmetry. In these examples, horizontal entropy gradients are presumably also accompanied by horizontal compositional gradients, which prompts the question of how fingering instabilities are modified in their presence. 

This question was in fact already addressed and mostly solved (at least in the linear regime) by \citet{holyer83}, who was interested in its application to mixing in the Earth's oceans -- the original thermohaline case.
In this letter, we apply in Section 2 the linear stability analysis of \citet{holyer83} to determine what effect horizontal gradients may have on double-diffusive mixing in stars. In Section 3 we carry out 3D DNS experiments of the process to study the nonlinear evolution of the instability, and the resulting turbulent mixing. 

\section{Model setup and linear stability analysis}

We build on the local model for fingering convection used by \citet{Traxleral2011}, and \citet{Brownal2013}, and include the effects of horizontal gradients as in the work of \citet{holyer83}. 
We model a small region within a star using a local Cartesian coordinate system, with gravity defining the vertical direction: ${\bf g} = -g{\bf e}_z$.  We use the Boussinesq approximation \citep{spiegelveronis1960}, which is valid as long as the domain height is much smaller than the local scaleheight. 

The background temperature and composition profiles are given by $T_0(x,z) = T_0 + T_{0z}z + T_{0x} x$ and $\mu_0(x,z) = \mu_0 +  \mu_{0z}z + \mu_{0x} x$, where the local means $T_0$, $\mu_0$, and the local gradients $T_{0z}$, $T_{0x}$, $\mu_{0z}$ and $\mu_{0x}$ are all constant. In addition, we assume that the local adiabatic temperature gradient, $T_{0z}^{\rm ad}$, is constant. The total fields $T_{\rm tot}(x,y,z,t)$ and $\mu_{\rm tot}(x,y,z,t)$ are then defined as $T_{\rm tot}(x,y,z,t) = T_0(x,z) + \tilde{T}(x,y,z,t)$ and $\mu_{\rm tot}(x,y,z,t) = \mu_0(x,z) + \tilde{\mu}(x,y,z,t)$.

For simplicity and consistency with the oceanographic case, we define the {\it potential temperature} as the temperature in excess of a theoretical adiabatic stratification: 
$\theta_{\rm tot} \equiv T_{\rm tot}- z T_{0z}^{\rm ad}$. Defining
\begin{equation}
\theta_{0z} = T_{0z}-T^{\rm ad}_{0z} \mbox{  and } \theta_{0x} = T_{0x}   \mbox{  ,}
\end{equation}
and $\theta_{\rm tot} = \theta_0(x,z) + \tilde{\theta}(x,y,z,t)$ as before, we then have  $\tilde{T} = \tilde{\theta}$. Potential temperature plays a role analogous to that of entropy in the Boussinesq approximation.

Density perturbations $\tilde \rho$ are related to $\tilde \theta$ and $\tilde \mu$ through the linearized equation of state: 
\begin{equation}
\frac{\tilde \rho}{\rho_0} = - \alpha \tilde \theta + \beta \tilde \mu \mbox{   , }
\label{eq:eos}
\end{equation}
where $\alpha = -(1/\rho_0)(\partial \rho/\partial T)$ and $\beta = (1/\rho_0)(\partial \rho/\partial \mu)$ are the coefficients of thermal expansion and compositional contraction. Following \citet{holyer83}, we assume that there is no net horizontal density gradient, which implies 
\begin{equation}
\alpha \theta_{0x} = \beta \mu_{0x} \mbox{  . }
\end{equation}
So-called ``density-compensating" horizontal gradients are common in the ocean \citep[e.g.][]{StommelFedorov67}. Presumably, non-compensating gradients would drive fast horizontal flows that rapidly equalize density along isobars, turning the system into a density-compensated one. We assume here that similar arguments apply to stellar interiors. 

When non-dimensionalized in as \citet{Traxleral2011} based on: (1) the expected finger width, so  $ [{\rm length}]= d = \alpha g \theta_{0z} /\kappa_T \nu$, where $\kappa_T$ is the thermal diffusivity and $\nu$ is the viscosity, (2) the thermal diffusion timescale across a finger, so  $[{\rm time}] = d^2/\kappa_T$, and (3) with $[\theta] = d \theta_{0z}$ and $[\mu] = \alpha d \theta_{0z}/\beta $, the governing equations are: 
\begin{eqnarray}
&& \frac{1}{\rm Pr} \left( \frac{\partial \bu}{\partial t} + \bu \cdot \nabla \bu \right) = - \nabla p + (\theta -\mu) \be_z + \nabla^2 \bu \mbox{   , } \nonumber \\ 
&& \frac{\partial \theta}{\partial t} + \bu \cdot \nabla \theta + s u + w  =  \nabla^2 \theta  \mbox{   , } \nonumber \\ 
&& \frac{\partial \mu}{\partial t} + \bu \cdot \nabla \mu + s u + \frac{w}{R_0}  = \tau \nabla^2 \mu \mbox{   , } \nonumber \\ 
&& \nabla \cdot \bu = 0 \mbox{   , }
\label{eq:goveqs}
\end{eqnarray}
where $\bu = (u,v,w)$ is the non-dimensional velocity field, and where for simplicity from here on, $p$, $\theta $ and $\mu$ represent the non-dimensional pressure, (potential) temperature and compositional perturbations away from the background. 

Four non-dimensional numbers appear: the Prandtl number ${\rm Pr} = \nu/\kappa_T$, the diffusivity ratio $\tau = \kappa_\mu/\kappa_T$ (where $\kappa_\mu$ is the compositional diffusivity), and finally 
\begin{equation}
R_0 = \frac{\alpha \theta_{0z} }{\beta \mu_{0z}}  \mbox{  and } s = \frac{\theta_{0x}}{\theta_{0z}} \mbox{   .}
\end{equation}
$R_0$ is the density ratio, which measures the strength of the stabilizing vertical potential temperature (entropy) gradient compared with the destabilizing $\mu$-gradient. Meanwhile, $-s$ is the slope of constant background $\theta$ surfaces. The slope of planes of constant (background) $\mu$ is then equal to $-s R_0$. 

In stellar interiors, ${\rm Pr}$ and $\tau$ can vary from $10^{-5}$ down to $10^{-11}$, and ${\rm Pr}$ is typically slightly greater than $\tau$. The slope $|s|$ could reasonably be assumed to be between $10^{-8}$ and $10^{-2}$, depending on the mechanism that generates the horizontal gradients. Finally, $R_0$ can take any value between one (near standard convective regions) and infinity, provided the region supports an inverse $\mu$-gradient. It is important to note that, as a result, the slope $|s| R_0$ of the constant-$\mu$ planes could in theory either be much smaller or much larger than one. However, we do not believe that the case $|s| R_0 >1$ is physically realistic in stars, and therefore always assume in what follows that $|s| R_0 \ll 1$. 

\citet{bainesgill1969} showed that in the absence of horizontal gradients, a system is only unstable to fingering convection if $1 < R_0 < \tau^{-1}$. Cases where $R_0 < 1$ are unstable to the Ledoux criterion, and therefore subject to dynamical convection. Cases where $R_0 > \tau^{-1}$ are completely stable to fingering. \citet{holyer83} showed that horizontal gradients always act to destabilize the system. In stellar interiors, where ${\rm Pr}, \tau \ll 1$, this statement applies regardless of the value of the density ratio. To see this, we seek normal form solutions of (\ref{eq:goveqs}) of the kind $q(x,y,z,t) = \hat q \exp(i lx + imy + ikz + \lambda t)$. Looking for the most unstable modes of this system can be done without loss of generality assuming that $m = 0$ \citep{holyer83}. This yields a cubic equation for the growth rate $\lambda$, of the form $\lambda^3 + a_2 \lambda^2 + a_1 \lambda + a_0 = 0$ with
\begin{eqnarray}
&& a_2 = |{\bf k}|^2(1+{\rm Pr}+\tau)  \mbox{   , } \\
&& a_1 = |{\bf k}|^4(\tau {\rm Pr} +\tau +{\rm Pr}) + {\rm Pr}\frac{l^2}{|{\bf k}|^2}\left(1-\frac{1}{R_0}\right) \mbox{   , }\nonumber \\
&& a_0 =|{\bf k}|^6\tau {\rm Pr} + {\rm Pr}\left[l^2(\tau - \frac{1}{R_0}) + kls(1-\tau)\right] \mbox{   , }\nonumber
\end{eqnarray} 
where ${\bf k} = (l,0,k)$ and $|{\bf k}|^2 = l^2 + k^2$. 

\begin{figure}[h]
\begin{center}
\includegraphics[width=0.45\textwidth]{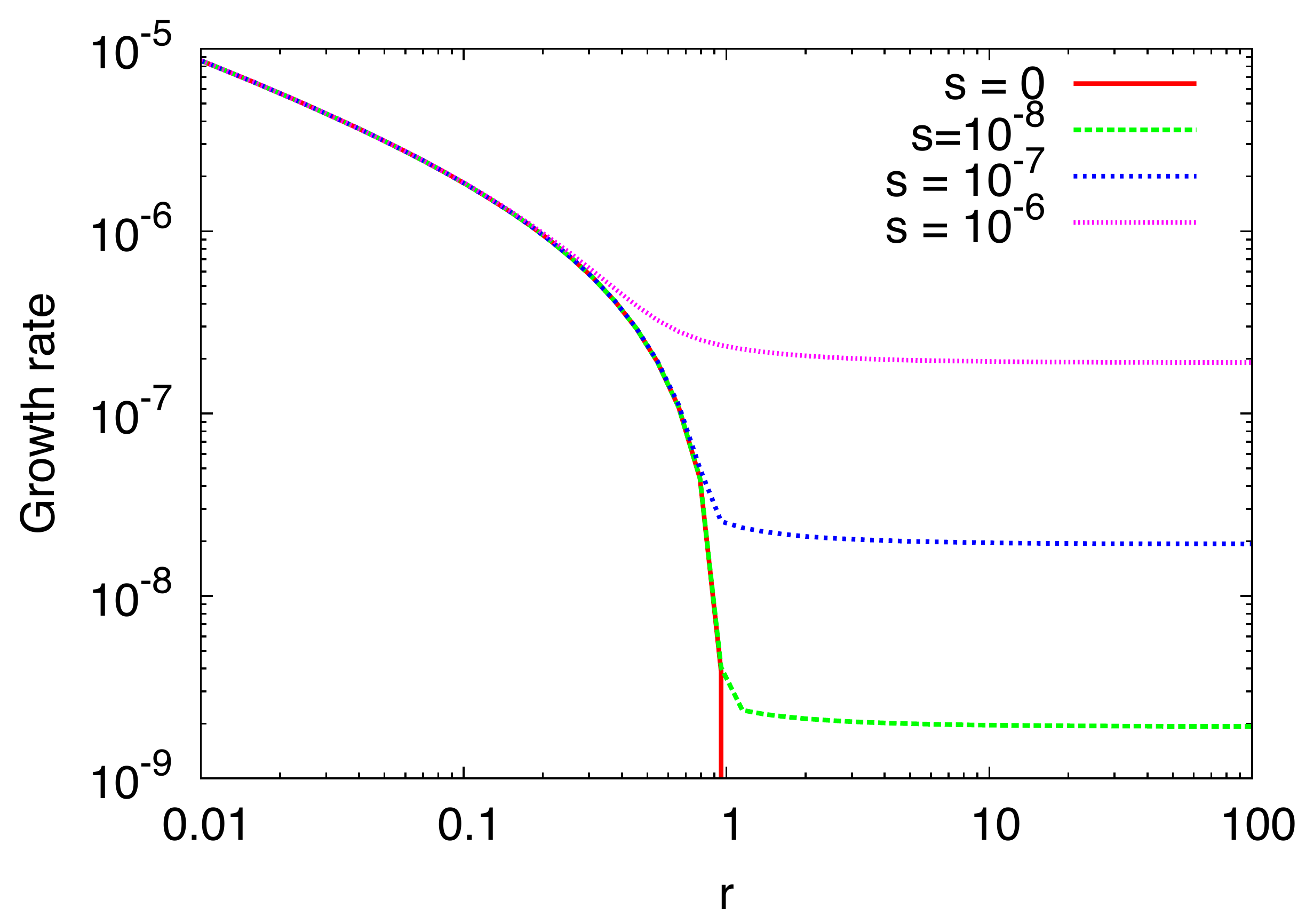}
\caption{Growth rates of the fastest growing mode as a function of $r$ (see Equation \ref{eq:littler}), for ${\rm Pr}  = \tau = 10^{-6}$, for various values of $s$. Note that the condition $sR_0 \ll 1$ for $R_0 > 1/\tau$ (the fingering-stable regime) implies that we must limit our study to $s \ll \tau$. 
}
\label{growthrates}
\end{center}
\end{figure}

Figure \ref{growthrates} illustrates the behavior of the growth rate of the fastest growing mode, that is, the solution of the aforementioned cubic maximized over all values of $k$ and $l$, as a function of the reduced density ratio $r$ defined as
\begin{equation}
r = \frac{R_0 -1}{\tau^{-1} -1} \mbox{   . }
\label{eq:littler}
\end{equation}

The case $s=0$ corresponds to the standard fingering instability, and illustrates how the unstable range is limited to $R_0 \in [1,\tau^{-1}]$, which corresponds to $r \in [0,1]$. However, we also find that unstable modes exist for $r >1$ as soon as $|s| > 0$.  Secondly, we see that the growth rate of the fastest-growing mode for $r \gg 1$ reaches a finite asymptotic value that scales with $s$. In fact, it can be shown using asymptotic analysis that for $r \rightarrow \infty $, for ${\rm Pr}$ and $\tau$ small and of the same order, and for $s \ll \tau$, the fastest-growing modes have the following properties: $\lambda$ is real, $k$ and $l$ have opposite signs, and 
\begin{equation}
\lambda \simeq \lambda_0\left( \frac{\tau}{{\rm Pr}} \right) s \mbox{   , } k \simeq - k_0\left( \frac{\tau}{{\rm Pr}} \right) \left( \frac{s}{\tau}  \right)^{1/2} \mbox{  and } l \simeq  l_0\left( \frac{\tau}{{\rm Pr}} \right) \left( \frac{s}{\tau}  \right)^{3/2}
\label{eq:lambdascal}
\end{equation}
where the functions $\lambda_0(\tau/{\rm Pr})$, $k_0(\tau/{\rm Pr})$ and $l_0(\tau/{\rm Pr})$ are of order unity (for $\tau/{\rm Pr}$ of order unity), and are related to one another. The actual expressions for $\lambda_0$, $l_0$ and $k_0$ are not particularly illuminating for our present purposes, however.

 Since $s \ll \tau$, we see from Equation (\ref{eq:lambdascal}) that while $l/k^3$ remains constant as $s$ decreases, $l/k$ tends to 0. Given that the unstable modes can simply be viewed as fingers whose orientation is inclined at an angle $\phi  = -\tan^{-1}(l/k)$ with respect to the horizontal, we see that they become more and more inclined as $s / \tau \rightarrow 0$. Their width, which is reasonably well-approximated by $k^{-1}$, increases as $s/\tau$ decreases. 

\begin{figure}[h]
\begin{center}
\includegraphics[width=0.5\textwidth]{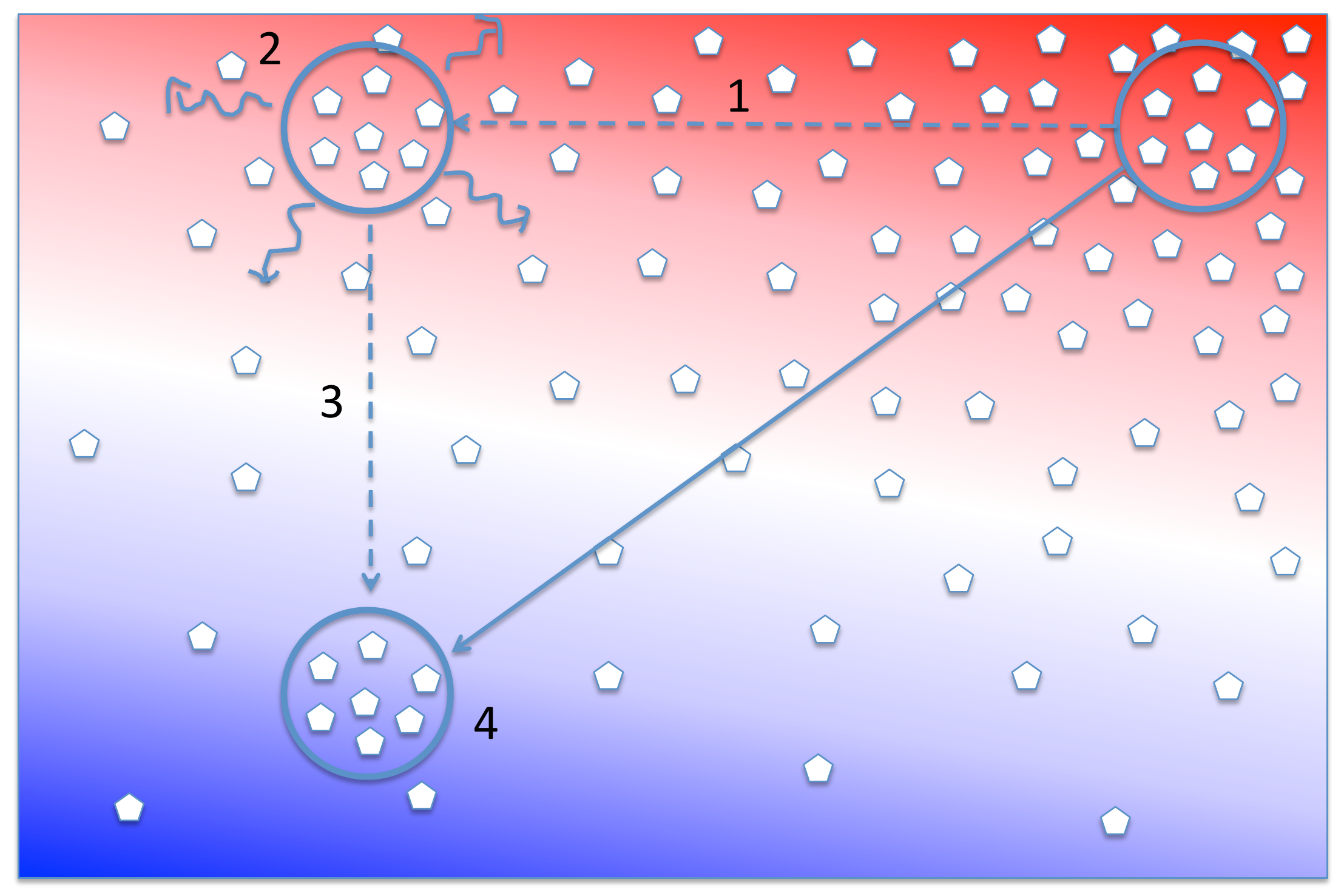}
\caption{Pictorial explanation of the basic instability mechanism (see main text for detail). The background potential temperature (entropy) stratification is represented by the colors, that of composition is represented by the density of symbols.}
\label{intrusive}
\end{center}
\end{figure}

Figure \ref{intrusive} explains the basic instability mechanism. 
Note that all angles are here greatly exaggerated, to illustrate the process more clearly. The background potential temperature and compositional fields are stratified both horizontally and vertically with $s>0$, leading to high-$\theta$/high-$\mu$ material in the top right of the figure, and low-$\theta$/low-$\mu$ material in the bottom left. Since density is constant on horizontal surfaces, any parcel of fluid can be displaced to the left without loss or gain of potential energy (1). Once displaced, the parcel rapidly loses heat to the surrounding fluid, but retains its higher $\mu$ content (2), as long as $\kappa_T > \kappa_\mu$. It becomes denser than the surroundings, and sinks (3). The entire process takes place in reality in a continuous fashion along inclined planes, as shown below in Section 3 (see Figure \ref{snapshots}), and causes a net transport of high-$\mu$ material from the top right to the bottom left of the domain (4). Note that similar steps (not shown), starting from the bottom left and moving a parcel first to the right, lead to an upward and rightward moving parcel. 


\section{Numerical simulations}

In order to study the nonlinear development of the system, and estimate the rate of turbulent mixing induced by the instability, we turn to 3D DNS of Equations (\ref{eq:goveqs}). We use the same code as in \citet{Traxleral2011b} and \citet{Brownal2013}, modified by adding the terms containing $s$ in the temperature and compositional equations. This code assumes that all perturbations are triply-periodic in space. We present the results of a single simulation, with parameters: ${\rm Pr }= \tau = 0.3$, $s = 0.1$, and $R_0= 5$. This corresponds to a reduced density ratio $r \simeq 1.5$, which would be stable to fingering convection if $s$ was 0.

At these parameter values, the fastest-growing mode has $\lambda \simeq 0.014$, $l \simeq \pm 0.02$ and $k \simeq \mp 0.20$. We therefore need a computational domain of horizontal length $L_x$ that is at least equal to $2\pi/l \simeq 315$ in order to contain it. In what follows, we choose $L_x \times L_y \times L_z = 500 \times 25 \times 500$. This domain is fairly thin in the $y$ direction to save on computational time, but nevertheless sufficiently thick to capture the 3D dynamics of the problem (see Garaud and Brummell, in preparation). The resolution used has $960 \times 48 \times 960$ effective grid-points.

In the oceanographic case, secondary instabilities form that tend to have even larger aspect ratios (smaller $l/k$) than the linear modes. Containing them in the computational domain becomes prohibitive. \citet{SimeonovStern2007} proposed that one can still model them numerically by using an {\it inclined} computational domain, with the tilt angle chosen so that the secondary modes are perfectly horizontal in the new reference frame. Preliminary simulations suggest that these modes at low ${\rm Pr}, \tau$ have the same angle as the primary modes. We thus tilt the domain by the angle $\phi = - \tan^{-1}(l/k) \simeq 5.7^{\circ}$ (see above) in the numerical simulations presented below. For clarity, any quantity measured in the tilted domain from here on is referred to with a prime (e.g. $w'$, $z'$, $u'$).

\begin{figure*}
\begin{center}
\includegraphics[width=\textwidth]{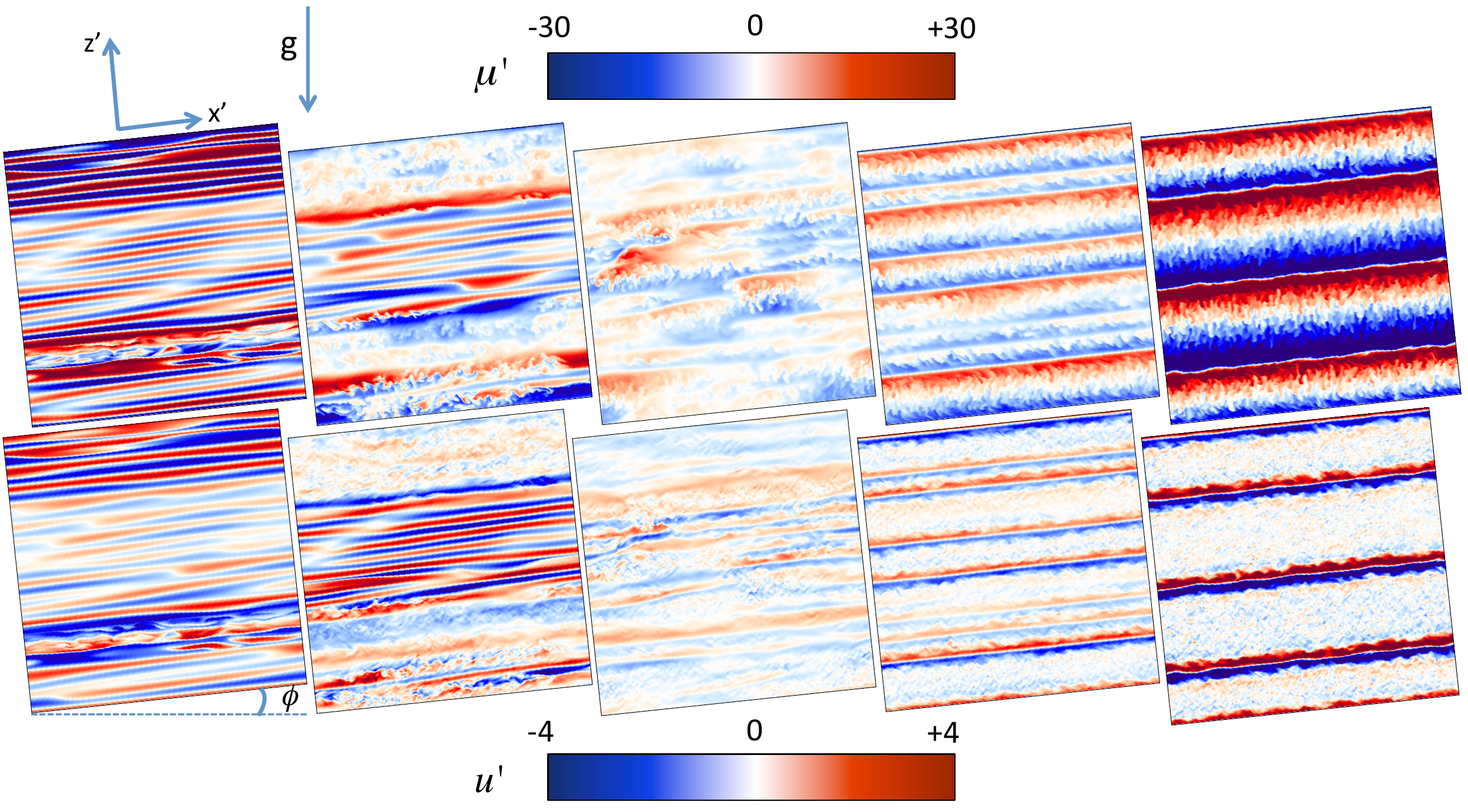}
\caption{Snapshots of the compositional perturbation $\mu'$ (top) and of the along-intrusion velocity $u'$ (bottom), at times (from left to right) $t = 1090$, $t = 1140$, $t = 1200$, $t=1710$, and $t=2800$.}
\label{snapshots}
\end{center}
\end{figure*}

Figure \ref{snapshots} shows successive snapshots of our simulation. At early times the behavior of the instability is as expected from linear theory. We find that the modes are horizontal in the inclined frame, and therefore inclined at an angle $\phi$ from the true horizontal. Alternating inclined fingers, whose width is roughly 30$d$, carry high-$\mu$ material downward, and low-$\mu$ material upward. Very rapidly, shear develops between the fingers and destabilizes them, locally mixing both entropy and composition. This weakens the stratification so that the local density ratio drops below $\tau^{-1}$, at which point regions that are properly fingering-unstable, separated by thin diffusive interfaces, appear. In the oceanographic context, these structures are called ``lateral intrusions" \citep[see the review by][Chapter 7]{Radko13}. The over-dense part of each intrusion propagates in this plot downward and to the left, while the under dense part propagates to the right and upward. Later, we see that the fingering region from one intrusion impinges on the one below, and the weakest of the two disappears. This results in a progressive coarsening of the system, which exhibits fewer and fewer yet gradually stronger intrusions with time.

\begin{figure}
\begin{center}
\includegraphics[width=0.5\textwidth]{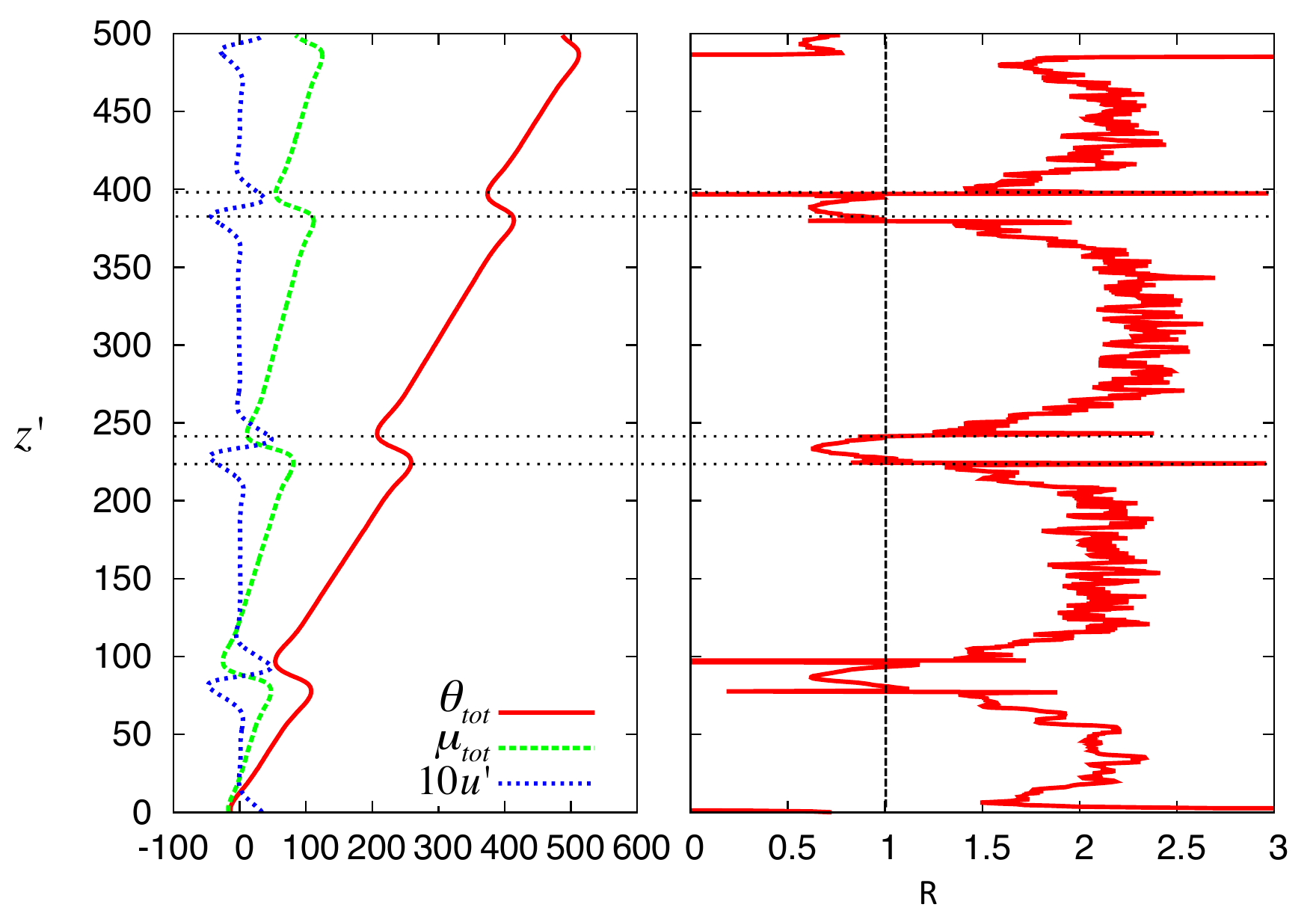}
\caption{Left: Variation with $z'$ of $\theta_{\rm tot}$ and $\mu_{\rm tot}$, and of $u'$, averaged along $x'$ and $y'$, at $t = 2800$. Right: Local density ratio $R(z')$, defined as the ratio of the $z'$-derivatives of the potential temperature and compositional profiles shown on the left. Horizontal lines highlight two of diffusive interfaces.}
\label{stratif}
\end{center}
\end{figure}

To better understand the structure of the intrusions, we show in Figure \ref{stratif} various profiles at time $t=2800$ (see final snapshot in Figure \ref{snapshots}). We clearly see extended regions where the total temperature and compositional fields increase with $z'$, which are associated with weak along-intrusion flows, and a local density ratio that remains smaller than $1/\tau$. These are the fingering layers observed in Figure \ref{snapshots}. Separating them are thin diffusive interfaces where both gradients reverse sign. These interfaces could in principle be semi-convectively unstable, in the sense that they support both an unstable potential temperature gradient, and a stable compositional gradient, and that their density ratio $R(z')$ lies above the critical value $({\rm Pr}+\tau)/({\rm Pr}+1) \simeq 0.46$ \citep{bainesgill1969}. However, the strong interfacial shear, combined with the fact that the region is quite thin, presumably stabilizes the interface against semi-convection (at least, in this simulation). 

The fingering regions cause substantial vertical transport, bringing high-$\theta$/high-$\mu$ material downward, and low-$\theta$/low-$\mu$ material upward. This net flux must be accompanied by an equivalent interfacial flux, since we do not observe any pile-up just above the interfaces. The interfacial flux is primarily caused by shear instabilities, as pure diffusion across the interface actually acts in the opposite direction (because of the reversal of the sign of both gradients). Our findings are generally consistent  with those in the oceanographic context \citep{SimeonovStern2007}.

\begin{figure}
\begin{center}
\includegraphics[width=0.5\textwidth]{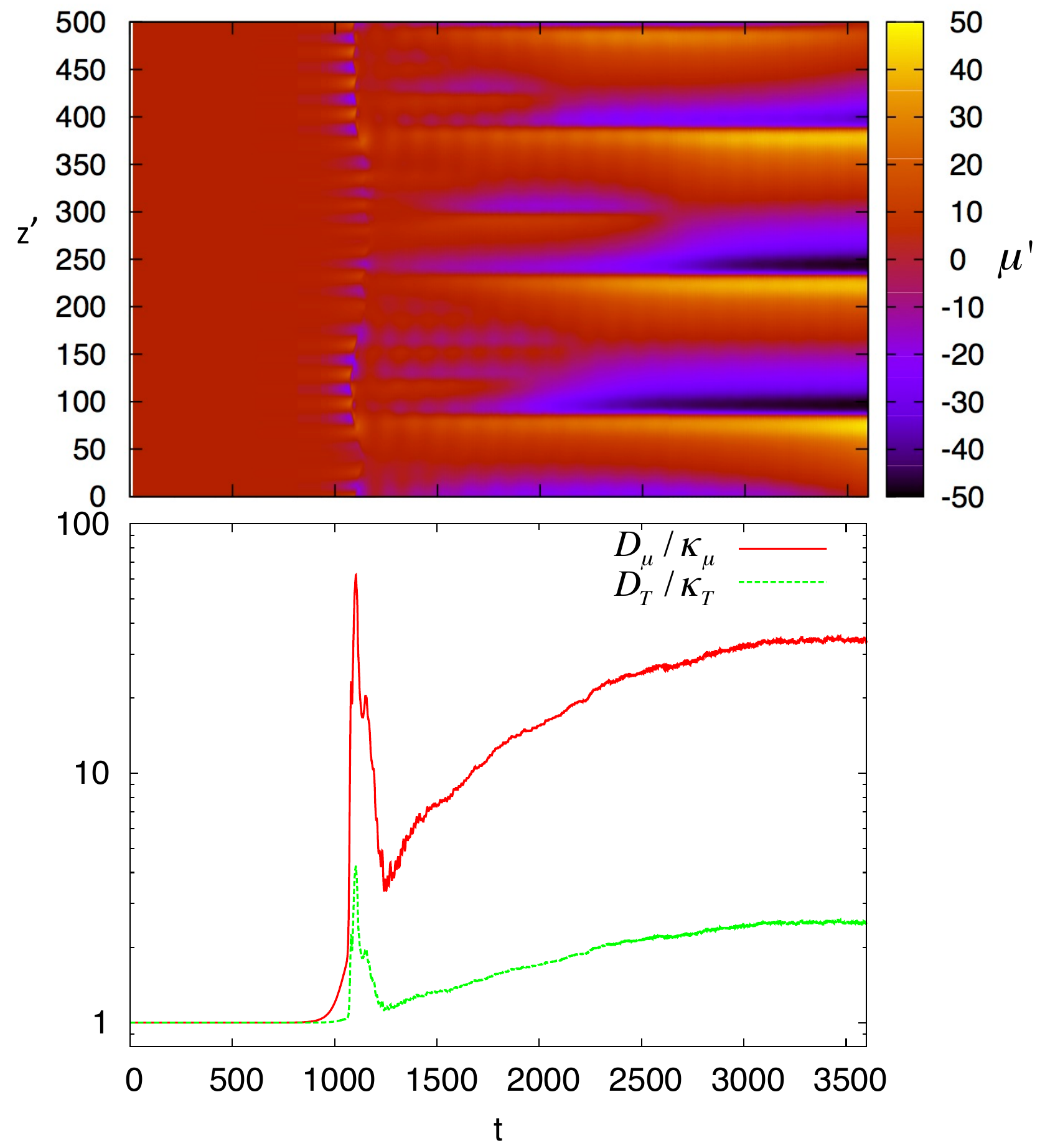}
\caption{Top: $x'$ and $y'$-averaged $\mu'$ profile as a function of time. Bottom: Non-dimensional effective diffusivities for vertical transport (see Equation \ref{eq:FtFmu}) as a function of time. }
\label{merger}
\end{center}
\end{figure}

Since $\phi$ is presumably very small in real stars, we can get an approximation of the amount of vertical transport in the system (i.e. along $z$) simply by considering the turbulent fluxes perpendicular to the intrusions (i.e. along $z'$). The resulting (dimensional) potential temperature and compositional fluxes, in the tilted domain, are then simply given by 
\begin{equation}
F_T \simeq - \kappa_T \theta_{0z} (1 - \langle w'\theta' \rangle) \equiv - D_T \theta_{0z} \mbox{  and  } F_\mu \simeq - \kappa_\mu \mu_{0z}  \left(1 - \frac{R_0}{\tau} \langle w'\mu' \rangle \right) \equiv - D_\mu \mu_{0z} \mbox{  ,} 
\label{eq:FtFmu}
\end{equation}
where the angular brackets denote a box-average. This defines the effective diffusivities $D_T$ and $D_\mu$. Figure \ref{merger} shows that the turbulent fluxes gradually increase as the intrusions coarsen. While $D_T/\kappa_T$ remains close to one, $D_\mu/\kappa_\mu$ is of the order of 30 at the end of the simulation, showing that compositional mixing is significantly enhanced by the intrusive instability, but that heat transport is not. Since we have not been able to follow the coarsening beyond the 3-intrusion phase, it remains to be seen whether the latter has stalled, or simply proceeds on a much longer timescale than we were able to integrate the simulation for.

\section{Conclusions and prospects}

We have shown that horizontal entropy and compositional gradients in a stably-stratified radiation zone can trigger new double-diffusive instabilities, that evolve to form stacks of fingering layers separated by sheared diffusive interfaces, called intrusions. Turbulent compositional mixing in this regime is significantly more efficient than pure diffusion, and could explain some of the long-standing discrepancies between theory and observations of stellar abundances, as in the case of low-mass RGB stars for instance \citep[e.g.][]{CharbonnelZahn07}.

Much work remains to be done, however. First and foremost, we need to extend our numerical analysis towards more realistic parameters (lower ${\rm Pr}$, $\tau$ and $s$), and carry out a comprehensive study of parameter space as in \citet{Brownal2013}. This is needed to determine how the cross-intrusion transport rates scale with governing parameters, and with the intrusion heights. This task is computationally expensive, since the already-large domain size and resolution need to be increased further when $s/\tau$, ${\rm Pr}$ or $\tau$ decrease. We also need to study the conditions under which the small-scale primary instability (see Section 2) transitions to the large-scale intrusive instability. While many models have been proposed to explain the formation of oceanic intrusions \citep[see, e.g.][]{walshruddick2000,SimeonovStern2007}, it remains unclear whether they apply in astrophysics. The unified mean-field theory developed by \citet{Traxleral2011b} is a promising approach in this respect. We also need to determine whether the intrusion coarsening process observed in our simulations proceeds indefinitely, or eventually stalls for a given intrusion height. Finally, we must determine whether any of the effects that were invoked as the possible origin of horizontal gradients (such as rotation, or slow large-scale flows) could have a sufficiently large influence on the development of both primary and secondary intrusive instabilities to invalidate our results. Nevertheless, these findings open an interesting set of new possibilities for mixing in stellar interiors, that had not been explored so far. 

\acknowledgements


\end{document}